\documentclass[%
 reprint,
 amsmath,amssymb,
 prd,
]{revtex4-2}

\usepackage{graphicx}
\usepackage{makecell}
\usepackage{tabularx}
\usepackage{dcolumn}
\usepackage{booktabs} 
\usepackage{multirow} 
\usepackage{siunitx}  
\usepackage{bm}
\usepackage[export]{adjustbox} 
\usepackage{array}
\usepackage[caption=false]{subfig}

\usepackage[usenames,dvipsnames,svgnames,table]{xcolor}
\usepackage{hyperref}

\definecolor{verdeoscuro}{rgb}{0, 0.5, 0}

\begin{document}

\preprint{APS/123-QED}

\title{Dark Matter Axion Detection with Neural Networks \\ at Ultra-Low Signal-to-Noise Ratio}



\author{Jos\'e Reina-Valero$^{a}$}
\email{Contact author: jose.reina@ific.uv.es}
\author{Alejandro D\'iaz-Morcillo$^b$}
\author{Jos\'e Gadea-Rodr\'iguez$^b$}
\author{Benito Gimeno$^a$}
\author{Antonio\:Jos\'e\:Lozano-Guerrero$^b$}
\author{Juan Monz\'o-Cabrera$^b$}
\author{Jose R. Navarro-Madrid$^b$}
\author{Juan Luis Pedreño-Molina$^b$}

\affiliation{
$^a$Instituto de Física Corpuscular (IFIC), CSIC-University of Valencia, Calle Catedr\'atico Jos\'e Beltr\'an Martínez, 2, 46980 Paterna (Valencia), Spain
}
\affiliation{
$^b$Departamento de Tecnolog\'ias de la Informaci\'on y las Comunicaciones,
Universidad Polit\'ecnica de Cartagena,
Plaza del Hospital 1, 30202 Cartagena, Spain
}


\date{\today}

\begin{abstract}
We present the first analysis of Dark Matter axion detection applying neural networks for the improvement of sensitivity. The main sources of thermal noise from a typical read-out chain are simulated, constituted by resonant and amplifier noises. With this purpose, an advanced modal method employed in electromagnetic modal analysis for the design of complex microwave circuits is applied. A feedforward neural network is used for a boolean decision (there is axion or only noise), and robust results are obtained: the neural network can improve by a factor of $5\cdot 10^{3}$ the integration time needed to reach a given signal to noise ratio. This could either significantly reduce measurement times or achieve better sensitivities with the same exposure durations.
\end{abstract}

\maketitle


\section{\label{sec:intro}Introduction}


The axion is probably the most elegant solution to the well-known strong CP (Charge-Parity) problem in the paradigm of QCD (Quantum ChromoDynamics). It was first postulated as the result of the SSB (Spontaneous Symmetry Breaking) of a new U(1) symmetry proposed by Robert Peccei and Helen Quinn \cite{Peccei_Quinn, Kim:2008hd} to explain the cancellation of the $\bar{\theta}$ angle. The Goldstone boson associated with this SSB is the axion (as pointed out by Weinberg \cite{Weinberg} and Wilczek \cite{Wilczek}). Due to the properties of this particle, it is the preferred candidate to be the main component of the Dark Matter in the Universe \cite{Willy_1,Willy_2,Willy_3}, reason why it has raised so much interest in recent years.

This axion has a feeble coupling to photons, implying that, by means of the inverse Primakoff effect \cite{Primakoff, Sikivie_2}, it can decay into photons under an external intense static magnetic field. If the generated photon is inside a resonant cavity and matches its frequency, it can be detected as a power excess. This setup, named haloscope, was first proposed by Sikivie \cite{Sikivie:1983ip} and aims to detect axions from the Milky Way halo.

Axion detection involves an important number of steps after the acquisition of a signal. Several power spectra must be acquired in an interval of time, which is usually in the range of minutes, hours or days per frequency point, depending on the experiment. The electronic background (EB) and its time variability must be removed from every spectrum and a grand unified spectrum is then constructed \cite{circuito_RADES}. The remaining systematics are removed using a Savitzky-Golay (SG) fit. Finally, the search for an axion is done by fitting its analytical form to the post-processed spectrum. Therefore, this kind of analysis is exhaustive and very high time and resources consuming. Additionally, this procedure must be carried out for each of the studied frequencies. In this article, we propose an alternative method that may be used in any axion detection experiment and in parallel to traditional techniques. The goal is to reduce the needed signal-to-noise ratio (SNR) through the application of a neural network to provide a reliable readout to detect the axion. Consequently, a drastic decrease in the time to determine whether an axion is present or not with a certain amount of probability could be obtained. Neural networks have raised a general interest in almost any branch of scientific knowledge, and are a powerful tool in pattern recognition problems. In this case, an exact and well-known full-wave modal method, the Boundary Integral-Resonant Mode Expansion 3D (BI-RME3D) \cite{bi-rme}, has been used to model the axion. Resonant noise has been added and the effect of the first low-noise amplifier (LNA) has been included in the analysis. Furthermore, a feedforward neural network (FNN) has been used as a predictive tool in order to evaluate the axion presence in a determined acquired signal. However, other machine learning (ML) techniques could have been employed with this purpose. In \cite{LACY2024115} a previous study for detecting a signal in the presence of noise is presented for applications in surveillance and remote sensing using a long short-term memory (LSTM) algorithm.
To the authors' knowledge, this is the first application of a neural network to the Dark Matter axion search problem aimed at enhancing the sensitivity.
\section{Proceedings and Methods}\label{sec:babyiaxo}

\begin{figure}
    \centering
    \includegraphics[scale = 0.55]{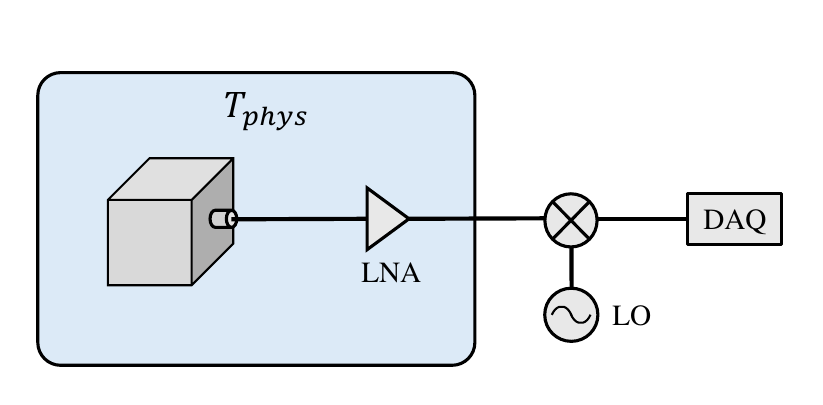}
    \caption{Readout chain employed for the axion detection study. On the left, the cavity is depicted along with the LNA amplifier at a temperature $T_{\mathrm{phys}}$. On the right, a pictorial representation of the receiver chain, constituted by a local oscillator and a DAQ system.}
    \label{fig:receiver_chain}
\end{figure}

A simplified receiver chain for the axion detection on which this study is based is depicted in Figure \ref{fig:receiver_chain}. The cavity and amplifier are positioned inside the cryogenic system, which has different temperature stages. Two system temperatures are considered for the numerical analysis: $T_{sys} = 1.2$ K and $T_{sys} = 4.0$ K, where $T_{sys}$ $=$ $T_{\mathrm{cav}}$ $+$ $T_{\mathrm{amp}}$ takes into account both the temperature of the cavity and the amplifier, respectively. The external magnetostatic field applied is $B_{e}$\:$=$\:10\:T. The values chosen are quite representative of a usual axion detection receiver chain, with more or less variation depending on the experiment.

The simulations performed in this work may include the aforementioned elements, since these are the ones mainly involved in hiding the axion RF (Radio-Frequency) signal under noise and systematic errors. After amplification, the signal is passed through a heterodyne receiver and down-converted to 0.1 MHz. Then, the signal is sampled in time-domain by the Data Acquisition (DAQ) system. 

A rectangular cavity has been employed, making use of the geometry of a standard rectangular waveguide, which is the WR-975, with dimensions $a$ $=$ $247.65$ mm (width), $b$ $=$ $123.825$ mm (height). The last dimension $d$ is fixed short-circuiting in the corresponding length. The considered axion mode will be the $\mathrm{TE}_{101}$ at a frequency of $\nu_{a} =$ 1\:GHz, implying that the cavity length is 
\begin{equation}
    d = \frac{1}{\sqrt{\left(\frac{2\,\nu_{a}}{c}\right)^2 - \left(\frac{1}{a}\right)^2}} = 188.31\, \mathrm{mm},
\end{equation}
where $\nu_{a}$ is the axion frequency and $c$ is the speed of light in vacuum. The assumed frequency of the axion has been chosen to be 1 GHz, since the low-GHz range has been extensively studied by several experiments, as reported in the technical literature \cite{CAPP_1, CAPP_2, CAPP_3, CAPP_4, HAYSTAC, RBF, TASEH, UF}. In order to simulate the scattering parameters of the cavity, CST Studio Suite \cite{CSTStudioSuite} has been employed. The unloaded quality factor of the cavity is $Q_{0}$ $=$ $1.1 \cdot 10^{5}$ for a cryogenic electrical conductivity $\sigma$ $=$ $1 \cdot 10^{9}$ S/m. A $50$ $\Omega$ coaxial probe has been connected to the cavity with a relative permittivity of $\varepsilon_{r} \, =\,  1$. The probe was introduced with a determined length in order to work in a critical coupling regime, in which half of the RF electromagnetic energy generated within the cavity is consumed while the remaining half is extracted for detection.

As mentioned previously, the photon produced by the axion decay is assumed to be $\nu_{a}$ $=$ $1$ GHz, and the KSVZ model has been addressed \cite{KSVZ_1, KSVZ_2}, implying that the coupling constant of the axion-photon interaction adopts the value $g_{a\gamma\gamma}$ $=$ $1.57 \cdot 10^{-15}$ $\mathrm{GeV}^{-1}$. Axion field amplitude $a_{0}$, defined as \cite{coreanos}
\begin{equation}
    a_{0} \approx \sqrt{\frac{2\rho_{\mathrm{DM}}\hbar^{3}/c}{m_{a}}},
\end{equation}
where $\hbar$ $=$ $h/\left(2\pi\right)$ is the reduced Planck constant; $a_{0}$ is the axion field amplitude, and adopts the value of $a_{0} \approx  6.36 \cdot 10^{-7}\, \mathrm{GeV}$ for a dark matter density $\rho_{\mathrm{DM}}$ $=$  $0.4\ \mathrm{GeV/cm^3}$ and an axion mass $m_{a} = 4.13$ $\mu$eV, where the KSVZ axion is assumed to constitute the whole percentage of Dark Matter in the Universe. In addition, the bandwidth of the axion has been introduced in the simulations, employing the well-known expression of the probability density function for the axion frequencies when introducing both the Maxwell-Boltzmann distribution and the laboratory reference frame movement \cite{Maxwell_Boltzman_distrib, HAYSTAC_analisis}:
\begin{equation}
\begin{aligned}
    f\left(\nu\right) = \ & \frac{2}{\sqrt{\pi}}\left(\sqrt{\frac{3}{2}}\frac{1}{r}\frac{1}{\nu_{a}\langle\beta^2\rangle}\right)\,\sinh\left(3r\sqrt{\frac{2\left(\nu - \nu_{a}\right)}{\nu_{a}\langle \beta^2\rangle}}\right) \\
    & \times \mathrm{exp}\left(-\frac{3}{2}r^2\, -\, \frac{3\left(\nu - \nu_{a}\right)}{\nu_{a}\langle\beta^2\rangle}\right),
\end{aligned}
\end{equation}
where $\langle\beta^2\rangle$ $=$ $\langle v^2\rangle/c^2$, being $\langle v^2\rangle$ $=$ $3\,k_{B}T/m_{a}$ the axion root mean square velocity (where $k_{B}$ is the Boltzmann constant and $T$ is the temperature of the virialized axion); and $r$ $=$ $v_{lab}/\sqrt{\langle v^2\rangle}$ is the ratio between the laboratory reference frame velocity $v_{lab}$ and $\sqrt{\langle v^2\rangle}$. This distribution endows the axion with a bandwidth of $\Delta\nu_{a}$\:$\approx 5$\:kHz, providing the axion with its well-known quality factor of $Q_{a}$ $=$ $10^{6}$.

\begin{figure}[t]
    \centering
    \includegraphics[scale = 0.45]{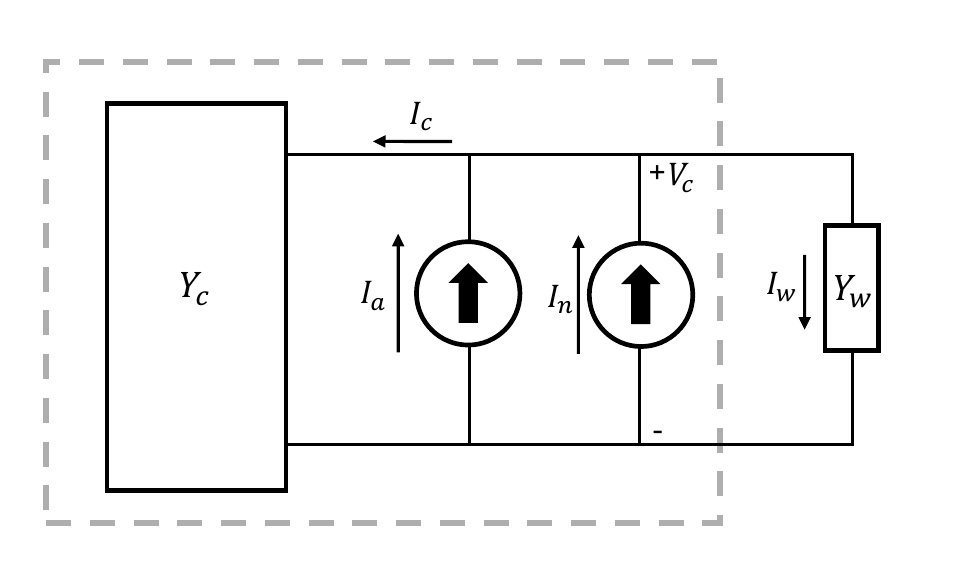}
    \caption{Equivalent circuit that shows how to add the axion and the resonant noise in the BI-RME3D formalism. Here, the current generators are both the axion, $I_{a}$, and the intrinsic noise of the cavity, $I_{n}$. $I_{w}$ is the current delivered to the coupled port (for detection) and $I_{c}$ is the current that will be eventually consumed by the cavity as ohmic losses. $Y_{c}$ represents the admittance of the one port cavity, and $V_{c}$ is the voltage detected in the coaxial port. Note that all the magnitudes previously defined are complex phasors, emphasizing that the BI-RME3D technique provides information about both magnitude and phase of the RF detected signal.}
    \label{fig:circuito_BI-RME}
\end{figure}

In order to analyze the RF voltage extracted from the cavity ($V_{c}$ in Figure \ref{fig:circuito_BI-RME}), the advanced modal method  is applied. This method provides a wide-band full-wave model which was developed during the eighties and nineties of the last century in the Università degli Studio di Pavia (Italy). It has been properly used for obtaining the structure of the electromagnetic fields inside an arbitrarily-shaped cavity with a given number of ports connected to it, and also for the efficient and accurate design of complex microwave passive components. Since the mathematical formulation of the method has been described in several publications \cite{advanced_modal_analysis, bi-rme, bi-rme_Pablo}, the main conclusions are outlined in this work: this method allows to establish a connection between the axion-photon coupling inside a microwave cavity and the classical electromagnetic network theory, as it can be seen in Figure \ref{fig:circuito_BI-RME}, considering the axion as a current source, $I_{a}$, which eventually divides into two parts, one delivered to the port (such as a coaxial or waveguide port), $I_{w}$, and the other one consumed by the cavity and dissipated as ohmic losses (Joule effect), $I_{c}$. Since there is no approximation during the formulation, the provided results are exact. However, since it is a modal method, the number of resonant modes $M$ in which results are truncated must be eventually chosen.

BI-RME3D provides, among others, the current generated by the axion $I_{a}$ and the voltage measured at the coaxial port $V_{c}$ as complex phasors, thus obtaining information of both amplitude and phase as a function of frequency. In addition, the extracted power $P_{w}$ can be calculated too. Resonant noise extracted from the cavity can also be added to the formulation as a random Gaussian-generated current phasor $I_{n}$, that is normalized to the expected noise power $k_{B}\,T_{\mathrm{cav}}\,\Delta\nu$. The expression of the total current from both the axion and the resonant noise is given by
\begin{widetext}
\begin{equation}
    \begin{aligned}
        I_{T} & =  -\sum_{m=1}^{M}F_{m1}^{(1)}\frac{\kappa_{m}}{\kappa_{m}^2 - k^2}\int_{V}\vec{E}_{m}\left(\vec{r}\ ^\prime\right)\cdot \left(\vec{J}_{a}\left(\vec{r}\ ^\prime\right) +  \vec{J}_{n}\right)\ dV^\prime =  \\
        & = -\sum_{m=1}^{M}F_{m1}^{(1)}\frac{\kappa_{m}}{\kappa_{m}^2 - k^2}\left(\int_{V}\vec{E}_{m}\left(\vec{r}\ ^\prime\right)\cdot \vec{J}_{a}\left(\vec{r}\ ^\prime\right)\ dV^\prime +  \int_{V}\vec{E}_{m}\left(\vec{r}\ ^\prime\right)\cdot \vec{J}_{n}\ dV^\prime\right) = I_{a} + I_{n},
    \end{aligned}
\end{equation}
\end{widetext}
where the summation is over the number of $M$ resonant cavity modes considered; $F_{m1}^{\left(1\right)}$ is the coupling integral between the magnetic field of the $m$-th normalized resonant cavity mode and the normalized magnetic field of mode 1 in port 1; $\kappa_{m}$ is the perturbed wavenumber of the $m$-th normalized resonant cavity mode taking into account the ohmic losses \cite{bi-rme_Pablo}, $k$ is the wavenumber corresponding to the frequency scan $k=\omega/c$, being $\omega=2\pi \nu$; $\vec{E}_{m}$ is the normalized electric field of the $m$-th normalized resonant cavity mode; and $\vec{J}_{a}$ and $\vec{J}_{n}$ are the axion and noise equivalent electric current densities, respectively.
The measured voltage at the cavity port is expressed as $V_{c} = (I_{a} + I_{n})/(Y_{w} + Y_{c})$, where $Y_{w}$ and $Y_{c}$ are the waveguide and cavity admittances, respectively. In this way, the delivered power can be expressed as:
\begin{equation}
    P_{w} = \frac{\left|I_{a} + I_{n}\right|^2}{2\left|Y_{w} + Y_{c}\right|^2}\mathrm{Re}\left(Y_{w}^{*}\right).
\end{equation}

\begin{figure}
    \centering
    \includegraphics[scale = 0.5]{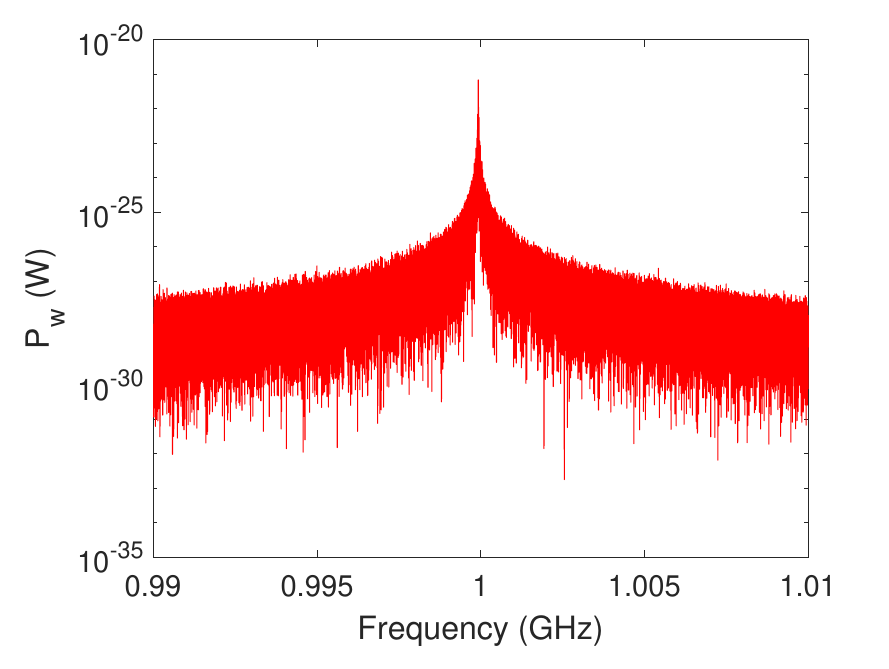}
    \caption{Resonant noise power extracted from the cavity at a temperature of $T_{\mathrm{cav}}$ $=$ $10$ mK.}
    \label{fig:resonant_noise}
\end{figure}

\begin{figure}
    \centering
    \includegraphics[scale = 0.55]{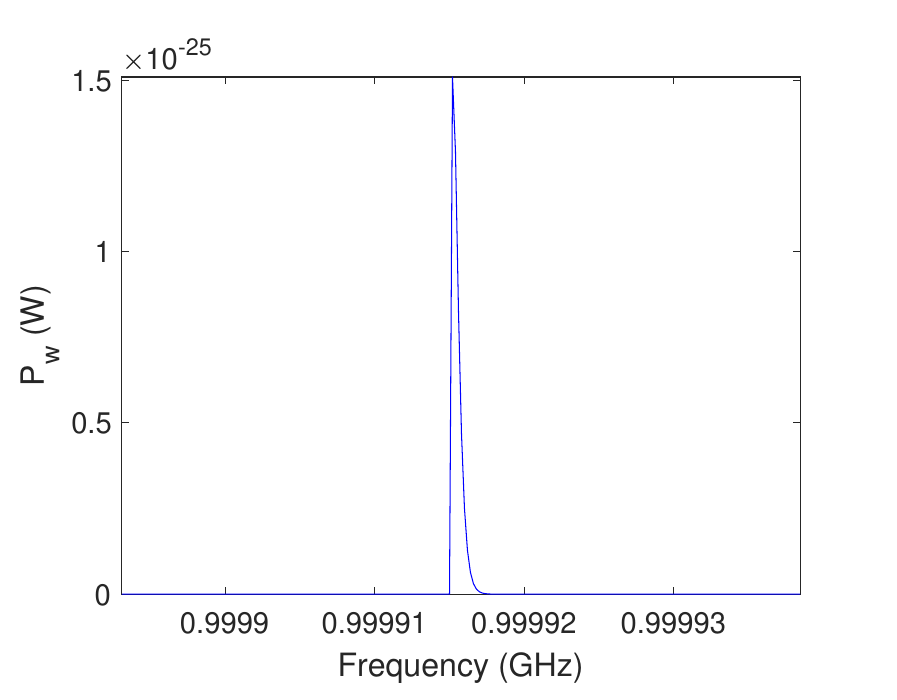}
    \caption{Power extracted from the cavity excited by the axion decay. The characteristic bandwidth of the axion introduced by the shifted Maxwell-Boltzmann distribution can be seen, endowing the axion with a quality factor of $Q_{a}$ $=$ $10^{6}$.}
    \label{fig:axion_spectrum}
\end{figure}

In Figure \ref{fig:resonant_noise} it is depicted the frequency sweep of the noise power extracted from the cavity (without axion) at a temperature of 10 mK, and in Figure \ref{fig:axion_spectrum} the power extracted from the cavity excited by the axion decay is plotted.

The signal-to-noise ratio is a common measurement to compare the power level between both the axion and the noise. It is defined, through the Dicke radiometer equation \cite{Dicke:1946glx}, as:
\begin{equation}
    \mathrm{SNR} = \frac{P_{w}}{k_{B}T_{sys}}\sqrt{\frac{t}{\Delta\nu}},
\end{equation}
where $t$ is the exposure time of the experiment. In order to maximise the signal-to-noise ratio, axion bandwidth $\Delta \nu_{a}$ has been considered as the detection bandwidth.

Referring to the neural network, a feedforward neural network has been employed, with only one neuron in the hidden layer. Since the output from the neural network is just boolean (i.e., there is axion or not), more neurons would produce overfitting in the results. More complex types of neural networks could be applied, but this task is left for future prospects of this work. However, the employment of a feedforward neural network aligns properly with the scope of this work, which is to show that even a simple kind of neural network can have significant impact in the analysis procedure.

\section{Numerical Results}

\begin{figure}
    \centering
    \includegraphics[scale = 0.5]{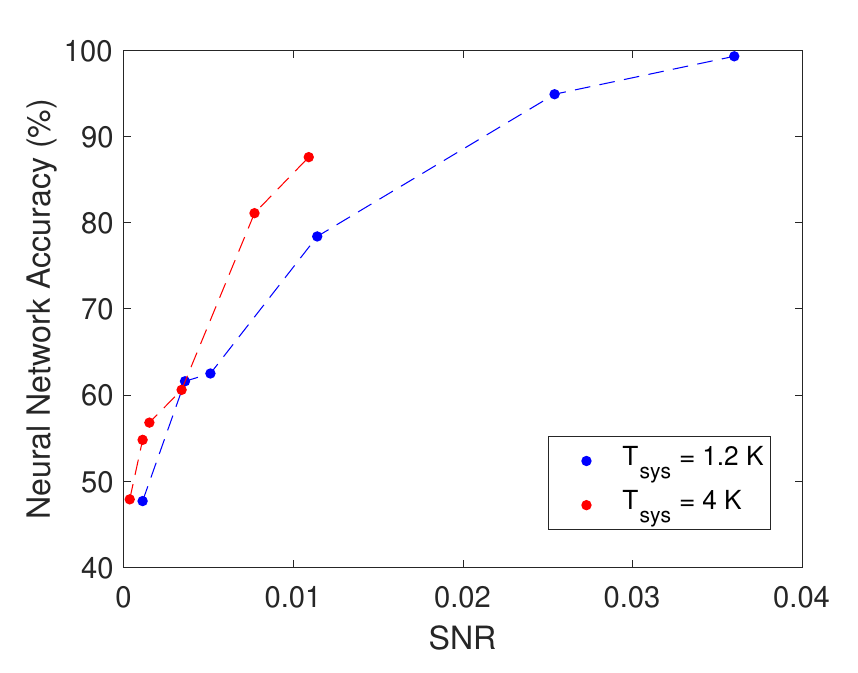}
    \caption{Neural network accuracy for given values of SNR. Each curve makes reference to a value of the system noise temperature.}
    \label{fig:Accuracy_vs_SNR}
\end{figure}

Once the methods are described, the neural network is applied to simulations of the axion RF voltage $V_{c}$ as well as resonant and amplifier noises. Relative to the training procedure, the neural network learns from 5000 trainings for each SNR. A random number is generated in order to decide if the RF signal recorded has axion or only thermal noise. 1000 tests were performed after the learning procedure, and the precision of the neural network is calculated as the ratio of successful cases to the total number of cases. The results, shown in Figure \ref{fig:Accuracy_vs_SNR} and Table \ref{tab:accuracy_vs_nbatch}, demonstrate reliable performance in identifying the axion RF signal. These findings align closely with those reported in \cite{LACY2024115}. By averages, it is understood that we refer to the number of signal time intervals added and subsequently averaged for the decrease of the noise standard deviation. 

\begin{figure}
    \centering
    \includegraphics[scale = 0.5]{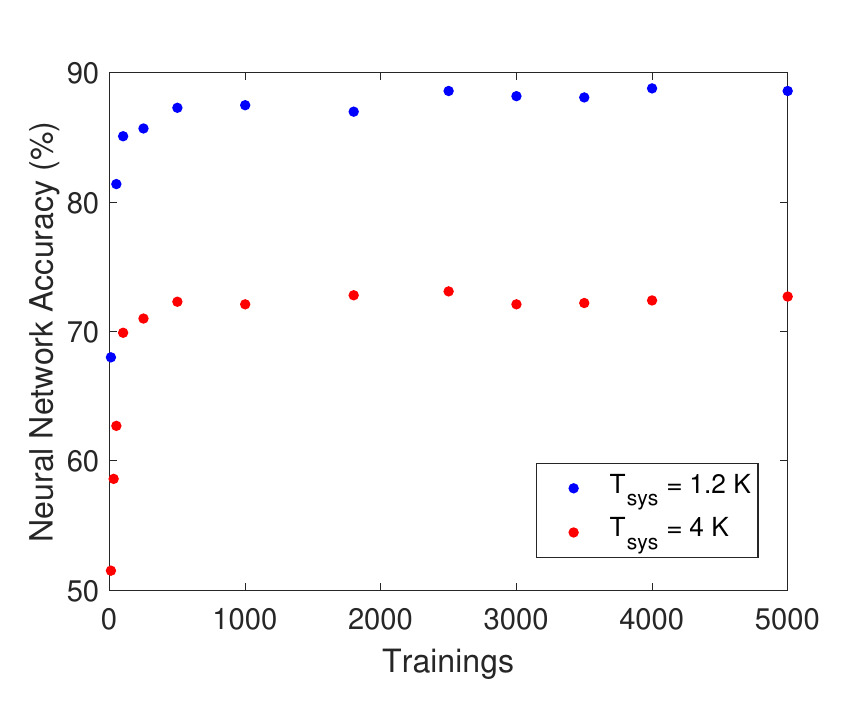}
    \caption{Accuracy improvement for different number of trainings depending on the system noise temperature. 250 averages were considered for obtaining the data.}
    \label{fig:Accuracy_vs_ntrain}
\end{figure}

In addition to the mentioned calculations, several trainings have been considered for a fixed number of averages (250 in our case) with the aim of observing the neural network performance when varying the trainings introduced (see Fig. \ref{fig:Accuracy_vs_ntrain}), showing the expected behaviour of rapid increase followed by saturation.


\begin{table*}[ht]
\centering
\begin{tabular}{c | cccc | cccc }
\toprule
 & \multicolumn{4}{c|}{$T_{sys} = 1.2$ K} & \multicolumn{4}{c|}{$T_{sys} = 4.0$ K}\\
\cmidrule(lr){2-5} \cmidrule(lr){6-9}
Averages & \multicolumn{1}{c}{$\mathrm{SNR_0}$} & \multicolumn{1}{c}{Accuracy (\%)} & \multicolumn{1}{c}{$\mathrm{SNR_{equiv}}$} & \multicolumn{1}{c}{$\mathcal{T}$} & \multicolumn{1}{c}{$\mathrm{SNR_0}$} & \multicolumn{1}{c}{Accuracy (\%)} & \multicolumn{1}{c}{$\mathrm{SNR_{equiv}}$} & \multicolumn{1}{c}{$\mathcal{T}$}\\
\midrule
1    & 0.0011 & 47.7 & 0.639 & 3.4 $\cdot 10^{5}$ & 3.4 $\cdot$ $10^{-4}$ & 47.9 & 0.642 & 3.6 $\cdot 10^{6}$\\
10   & 0.0036 & 61.6 & 0.871 & 5.9 $\cdot 10^{4}$ & 0.0011 & 54.8 & 0.752 & 4.7 $\cdot 10^{5}$\\
20   & 0.0051 & 62.5 & 0.887 & 3.0 $\cdot 10^{4}$ & 0.0015 & 56.8 & 0.786 & 2.7 $\cdot 10^{5}$\\
100  & 0.0114 & 78.4 & 1.237 & 1.2 $\cdot 10^{4}$ & 0.0034 & 60.6 & 0.852 & 6.3 $\cdot 10^{4}$\\
500  & 0.0254 & 94.9 & 1.951 & 5.9 $\cdot 10^{3}$ & 0.0077 & 81.1 & 1.314 & 2.9 $\cdot 10^{4}$\\
1000 & 0.0360 & 99.3 & 2.697 & 5.6 $\cdot 10^{3}$ & 0.0109 & 87.6 & 1.538 & 2.0 $\cdot 10^{4}$\\
\bottomrule
\end{tabular}%
\caption{Accuracy of the neural network for different values of averages and system noise temperatures. The number of trains is 5000 and the number of tests is 1000. The equivalent SNR after the application of the neural network, $\mathrm{SNR}_{\mathrm{equiv}}$, and the time improvement, $\mathcal{T}$, are also shown.}
\label{tab:accuracy_vs_nbatch}
\end{table*}

To clearly assess the capability of the neural network, some values of interest are analyzed in Table \ref{tab:accuracy_vs_nbatch}, where time improvement $\mathcal{T}$ is calculated as the ratio between the time required to reach the equivalent SNR without the neural network, $t_{\mathrm{equiv}}$ , and the original SNR of the signal introduced to the neural network, $t_{0}$:
\begin{equation}
    \mathcal{T} = \frac{t_{\mathrm{equiv}}}{t_{0}} = \left(\frac{\mathrm{SNR}_{\mathrm{equiv}}}{\mathrm{SNR}_{0}}\right)^2,
\end{equation}
where $\mathrm{SNR}_{\mathrm{equiv}}$ is calculated as the number of standard deviations in a Gaussian distribution $\mathcal{N}\left(0,1\right)$ that correspond to the associated neural network accuracy.

A robust performance is observed in the results, obtaining remarkable values of $\mathcal{T}$. For instance, with a SNR $=$ $0.036$ the accuracy of the neural network is 99.3\%, close to an equivalent SNR of 3. This has strong consequences on the exposure time needed to reach a given signal-to-noise ratio: for instance, if an experiment requires 100 days of exposure time in order to reach a SNR = 3, the application of this neural network is able to reduce this time by a factor $5 \cdot 10^{3}$, allowing to reach the desired SNR in approximately half an hour. Moreover, for slightly higher SNR, the accuracy of the neural network would be close to 100\%. 

Referring to the application of this methodology to a real data-taking procedure, the question of how to train the neural network comes out: if the trainings are experimental measurements, no time is gained since the time saved in the data-taking is substituted by the time wasted in neural network training. Thus, the value of this methodology is to simulate, with a high degree of realism, the thermal noise and any possible systematic errors (aspect that is left for future prospects) that could appear in a hypothetical measurement, and to train the neural network with this simulations. This kind of simulation would require a more detailed description than the one made in this work. However, the goal of this letter is to point out the remarkable capacity of the neural network to distinguish the axion from the most relevant generators of thermal noise in a real readout chain.

\section{Conclusions}

A basic model of a neural network, consisting on a feed forward network with one neuron in the hidden layer, has been applied to the simulation of a haloscope experiment. 

The application of a neural network to improve the SNR in an axion detection experiment has been done for the first time in this work to the knowledge of authors, obtaining significant results that have a direct impact in the exposure time needed by the experiment in order to reach a given sensitivity. For a SNR $= 0.036$, the equivalent SNR obtained by applying the neural network is close to 3, a common criteria to determine the presence of a potential candidate. This implies a time improvement by a factor of approximately 5$\cdot 10^{3}$. This is a remarkable result, since this reduction in exposure time would allow not only to faster sweeps but also to reach better sensitivities maintaining the same exposure time as nowadays, allowing to probe a significant lower value of the axion-photon coupling constant $g_{a\gamma\gamma}$. Owing to the simplicity of this technique, it can be applied either as the primary method for distinguishing axions from the thermal background in any current axion detection experiment, or as a complementary tool alongside the standard statistical analyses mentioned earlier. However, it must be pointed out that the ideal scenario to take profit from this time saving is to train the neural network with very precise simulations of the thermal noise and systematic errors, allowing for a fast learn process of the neural network.

This SNR improvement could also be applied to another type of experiments trying to detect a different kind of phenomena which requires extremely low sensitivities. For instance, the detection of HFGWs (High-Frequency Gravitational Waves) with haloscopes nowadays needs a substantial improvement in order to reach the expected HFGW strains \cite{HFGW_Diego,FLASH,BabyIAXO_HFGW}, and one of the main troubles is the exposure time, being only able to accumulate signal while the HFGW is passing through the cavity, making it difficult to reach the desired strain. With the proposed technique in this work, the expected experimental sensitivity could suffer a strong increment by only applying a neural network to the data study. This opens a new line of research, not only for axions or HFGWs, but also for any kind of physical phenomena that generates a feeble signal.

\section{\label{acknowledgments} Acknowledgments}
This work is part of the R$\&$D project PID2022-137268NBC53 and PID2022-137268NA-C55, funded by MICIU/AEI/10.13039/501100011033/ and by “ERDF/EU”. J. Reina-Valero has the support of ``Plan de Recuperación, Transformación y Resiliencia (PRTR) 2022 (ASFAE/2022/013)", founded by Conselleria d'Innovació, Universitats, Ciència i Societat Digital from Generalitat Valenciana, and NextGenerationEU from European Union.

\bibliography{references}

\end{document}